\documentclass[aps,showpacs,onecolumn,superscriptaddress]{revtex4}
\usepackage[dvips]{graphicx}
\usepackage{amsmath}
\usepackage{color}

\begin{document}

\title{Thermal conductivities of one-dimensional anharmonic/nonlinear lattices: renormalized phonons and effective phonon theory}
\author{Nianbei Li}
\email{nbli@tongji.edu.cn} \affiliation{NUS-Tongji Center for
Phononics and Thermal Energy Science and Department of Physics,
Tongji University, 200092 Shanghai, People's Republic of China}

\author{Baowen Li}
\email{phononics@tongji.edu.cn} \affiliation{NUS-Tongji Center for
Phononics and Thermal Energy Science and Department of Physics,
Tongji University, 200092 Shanghai, People's Republic of China}
\affiliation{Department of Physics and Centre for Computational
Science and Engineering, National University of Singapore, Singapore
117546, Republic of Singapore} \affiliation{NUS Graduate School for
Integrative Sciences and Engineering, Singapore 117456, Republic of
Singapore}

\begin{abstract}
Heat transport in low-dimensional systems has attracted enormous
attention from both theoretical and experimental aspects due to its
significance to the perception of fundamental energy transport
theory and its potential applications in the emerging filed of
phononics: manipulating heat flow with electronic anologs. We
consider the heat conduction of one-dimensional nonlinear lattice
models. The energy carriers responsible for the heat transport have
been identified as the renormalized phonons. Within the framework of
renormalized phonons, a phenomenological theory, {\it effective
phonon theory}, has been developed to explain the heat transport in
general one-dimensional nonlinear lattices. With the help of
numerical simulations, it has been verified that this effective
phonon theory is able to predict the scaling exponents of
temperature-dependent thermal conductivities {\it quantitatively}
and {\it consistently}.
\end{abstract}
\date{\today}

\maketitle

\section{Introduction}
The Fourier's heat conduction law states that the heat current
flowing through a system is proportional to the temperature gradient
imposed on the two ends of the system: $j=-\kappa \nabla{T}$, where
$j$ is the heat current, $\nabla{T}$ is the temperature gradient and
the proportionality $\kappa$ is defined as the thermal conductivity.
A diffusive or normal heat conduction requires that $\kappa$ is
independent to the system length $L$ in the thermodynamical limit
$L\rightarrow\infty$. However, the discovery of anomalous heat
conduction \cite{liusha2012EPJB} that $\kappa\propto L^{\alpha}$
with $0<\alpha<1$ for one-dimensional Fermi-Pasta-Ulam $\beta$
(FPU-$\beta$) lattice has casted doubt on the validity of Fourier's
heat conduction law on low-dimensional systems and stimulated
intensive studies on this issue from both theoretical
\cite{bonetto2000,lepri2003PR377,libaowen2005CHAOS15,wangjiansheng2008EPJB62,dhar2008AP57}
and experimental \cite{chang2008PRL101} attempts. On the other hand,
it has been demonstrated that the nonlinearity can be utilized to
design novel nanoscale solid-state thermal devices such as thermal
diodes\cite{terrano2002PRL88,libaowen2004PRL93,libaowen2006PRL95,chang2006Science314},
thermal transistors\cite{libaowen2006APL88}, thermal logic
gates\cite{wanglei2007PRL99} and thermal
memories\cite{wanglei2008PRL101,xierongguo2011AFM21} which gives
birth to the innovating field of phononics: manipulating/controlling
heat flow and processing information with
phonons\cite{linianbei2012RMP}. The distinctive and unique transport
property of low-dimensional system has posted great challenge to the
complete microscopic transport theory. Therefore, any theoretical
attempt towards a thorough understanding of the heat transport in
general one-dimensional nonlinear lattice systems is timely and
highly desirable.

In order to reveal the physical mechanism underlying the heat
transport in one-dimensional nonlinear lattices, the energy carriers
responsible for heat transport must be identified on the first
place. Although the linear Harmonic lattice can only sustain the
vibrations of phonons, it has been known that there are more than
one type of excitation modes in nonlinear lattices, i.e. the
renormalized
phonons\cite{alabiso1995JSP79,alabiso2001JPA34,lepri1998PRE58,gershgorin2005PRL95,gershgorin2007PRE75,linianbei2006EPL75,linianbei2007EPL78,hedahai2008PRE78},
solitons\cite{wattis1993JPA26,friesecke1994CMP161,zhangfei2000PRE61,zhangfei2001PRE64}
and breathers\cite{flach1998PR295,flach2008PR467}. In particular,
the stable solitons which almost do not interact with each other has
been argued to be the origin of the anomalous heat conduction found
in FPU-$\beta$
lattice\cite{hubambi2000PRE61,aoki2001PRL86,zhaohong2005PRL94}. The
numerical calculations for the sound velocities of energy carriers
have also been found to follow the prediction of soliton
velocities\cite{aoki2001PRL86}. However, these calculations cannot
exclude the possibility of renormalized phonons as energy carriers
since the prediction of sound velocities of renormalized phonons is
not far from the prediction of solitons\cite{linianbei2006EPL75}. It
will be necessary to determine which excitation mode is the energy
carrier for nonlinear lattices and more accurate calculations need
to be performed to uncover this confusion\cite{linianbei2010PRL105}.

For the past decades, enormous efforts have been focused on the
study of normal or anomalous heat conduction, i.e. size-dependent
thermal conductivities of one-dimensional nonlinear
lattices\cite{liusha2012EPJB,bonetto2000,lepri2003PR377,libaowen2005CHAOS15,wangjiansheng2008EPJB62,dhar2008AP57}.
On the contrary, there are only a few works dealing with the
temperature-dependent behaviors of heat conduction
\cite{aoki2000PLA265,aoki2001PRL86,hubambi2005CHAOS15,lefevere2006JSMTE,linianbei2007EPL78,linianbei2007PRE76,hedahai2008PRE78,linianbei2009JPSJ78,nicolin2010PRE81,linianbei2011Pramana77}
which should be more relevant to experimental investigations. The
temperature dependence of heat conduction is caused by the
nonlinearity as the linear Harmonic lattice cannot display
temperature modulated response. Furthermore, the novel thermal
devices such as the thermal diodes are rooted on the fact of
temperature or nonlinearity modulated phonon spectrum. The
understanding of the temperature-dependent behavior of heat
conduction for nonlinear lattices is thus of both theoretical and
experimental importance. A transport theory based on the correct
energy carriers will definitely help us to unravel the underlying
physical mechanism for the heat transport in low-dimensional
nonlinear lattices.

The paper is organized as the follows: in Sec. II the general
one-dimensional nonlinear lattice models will be introduced and the
the concept of renormalized phonons will be discussed. The sound
velocities of energy carriers will be analyzed and determined
through detailed numerical calculations. Sec. III will then present
the phenomenological theory of effective phonon theory in the
framework of renormalized phonons. The predictions for
temperature-dependent thermal conductivities from effective phonon
theory will be compared with Non-Equilibrium Molecular Dynamics
(NEMD) simulation results. We will give conclusions and summaries in
Sec. IV.

\section{Nonlinear lattices and Renormalized Phonons}
The one-dimensional nonlinear lattice Hamiltonian has a general
form:
\begin{equation}
H=\sum^{N}_{i=1}\left[\frac{p^2_i}{2}+V(x_{i+1},x_i)+U(x_i)\right],
\end{equation}
where $p_i$ and $x_i$ denote the momentum and deviation from
equilibrium position for the $i$-th atom, respectively. For
simplicity, the periodic boundary condition $x_1\equiv x_{N+1}$ is
often applied where $N$ is the number of atoms or the size of
lattice in dimensionless unit since $L=Na$ where $a$ is the lattice
constant and can be set as unity\cite{linianbei2012RMP}. The
inter-atom potential $V(x_{i+1},x_i)$ and the on-site potential
$U(x_i)$ also assume the following general expressions:
\begin{equation}
V(x_{i+1},x_i)=\sum^{\infty}_{s=2}v_s\frac{\left(x_{i+1}-x_{i}\right)^s}{s},\,\,\,
U(x_i)=\sum^{\infty}_{s=2}u_s\frac{x^s_i}{s}
\end{equation}
For simplicity, the coefficients $v_s$ and $u_s$ only take two
values $0$ and $1$. The celebrated FPU-$\beta$ lattice
\begin{equation}\label{fpu-ham}
H=\sum^N_{i=1}\left[\frac{p^2_i}{2}+\frac{\left(x_{i+1}-x_i\right)^2}{2}+\frac{\left(x_{i+1}-x_i\right)^4}{4}\right]
\end{equation}
has two nonvanishing coefficients $v_2=v_4=1$ while the most studied
on-site $\phi^4$ lattice
\begin{equation}\label{phi4-ham}
H=\sum^N_{i=1}\left[\frac{p^2_i}{2}+\frac{\left(x_{i+1}-x_i\right)^2}{2}+\frac{x_i^4}{4}\right]
\end{equation}
has two nonvanishing coefficients with $v_2=u_4=1$.

For the Harmonic lattice, it is well-known that the Hamiltonian can
be decomposed into the energy sum of normal modes (phonons) under
the transformation $q_k=\sum^N_{i=1}S_{ki}x_i$ and
$p_k=\sum^N_{i=1}S_{ki}p_i$ where the matrix element
$S_{ki}=\frac{1}{\sqrt{N}}(\sin{2\pi ki/N}+\cos{2\pi ki/N})$. The
Hamiltonian in position space can thus be transformed into normal
mode space as
\begin{equation}
H=\sum^{N}_{i=1}\left[\frac{p^2_i}{2}+\frac{\left(x_{i+1}-x_i\right)^2}{2}\right]\rightarrow
H=\sum^{N}_{k=1}\left[\frac{p^2_k}{2}+\frac{\omega^2_{k}q^2_k}{2}\right],\,\,\omega_k=2\sin{\frac{\pi
k}{N}}\label{ham}
\end{equation}
where $\omega_k$ is the phonon frequency. The resulted total
Hamiltonian in phonon space is the energy summation of every
independent phonon modes. Therefore, the Harmonic lattice has no
phonon-phonon interaction and is called linear lattice in this
sense.

As for the nonlinear lattice, there does not exist any
transformation which can transfer the lattice Hamiltonian into the
sum of independent phonon modes. However, the nonlinear lattice
Hamiltonian can be expressed {\it approximately} as the energy
summation of phonon modes with renormalized frequencies due to the
nonlinearity.

The connection between the renormalized phonon frequency and the
nonlinearity originates from the generalized equipartition theorem:
\begin{equation}
k_{B}T=\left<q_k\frac{\partial{H}}{\partial{q_k}}\right>
\end{equation}
where $k_{B}$ is the Boltzmann constant and $\left<\cdot\right>$
denotes ensemble average. For Harmonic lattice, it is easy to see
from Eq. (\ref{ham}) that
\begin{equation}
k_{B}T=\omega^2_k\left<q^2_k\right>
\end{equation}
which is the familiar result that every degree of freedom (phonon
mode) shares the same amount of energy. For nonlinear lattice, a
similar relation holds for the renormalized phonon modes under the
mean-field approximation\cite{linianbei2006EPL75}:
\begin{equation}
k_{B}T\approx\hat{\omega}^2_k\left<q^2_k\right>
\end{equation}
where the renormalized phonon frequencies can be defined as
\begin{eqnarray}
\hat{\omega}_k&=&\sqrt{\alpha \cdot \omega^2_k +\gamma}\nonumber\\
\alpha&=&\sum^{\infty}_{s=2}v_s\frac{\left<\sum^N_{i=1}\left(x_{i+1}-x_i\right)^s\right>}{\left<\sum^N_{i=1}\left(x_{i+1}-x_i\right)^2\right>}\nonumber\\
\gamma&=&\sum^{\infty}_{s=2}u_s\frac{\left<\sum^N_{i=1}x_i^s\right>}{\left<\sum^N_{i=1}x_i^2\right>}\label{renorm}
\end{eqnarray}
and the renormalization coefficients $\alpha$ and $\gamma$ encode
the information of inter-atom potential and on-site potential,
respectively. For the simple Harmonic lattice with $v_2=1$, the
renormalization coefficient $\alpha=1$ and $\gamma=0$, the
renormalized phonon frequencies $\hat{\omega}_k=\omega_k$ thus
recovers the original phonon dispersion relation.

The nonlinear lattices can be classified into two categories: with
or without on-site potential. For lattices without on-site potential
as $U(x_i)=0$, we have $\gamma=0$ from Eq. (\ref{renorm}).
Therefore, the branch of renormalized phonon modes is acoustic-like
with
\begin{equation}\label{renorm-omega}
\hat{\omega}_k=\sqrt{\alpha}\cdot\omega_k
\end{equation}
For lattices with on-site potential as $U(x_i)\neq 0$, the
renormalization coefficient $\gamma\neq 0$ and the resulted phonon
branch is optic-like.

In order to understand the physical mechanism of the heat transport,
we need first to determine the energy carriers in nonlinear
lattices. Besides the renormalized phonons, other excitation modes
such as solitons and breathers existing in nonlinear lattices are
also possible candidates for energy carriers. One of the best way to
identify the energy carriers is to calculate the
temperature-dependent energy transport speeds and compare them with
the theoretical predictions for the sound velocities of different
excitation modes.

\begin{figure}
\includegraphics[width=0.8\columnwidth]{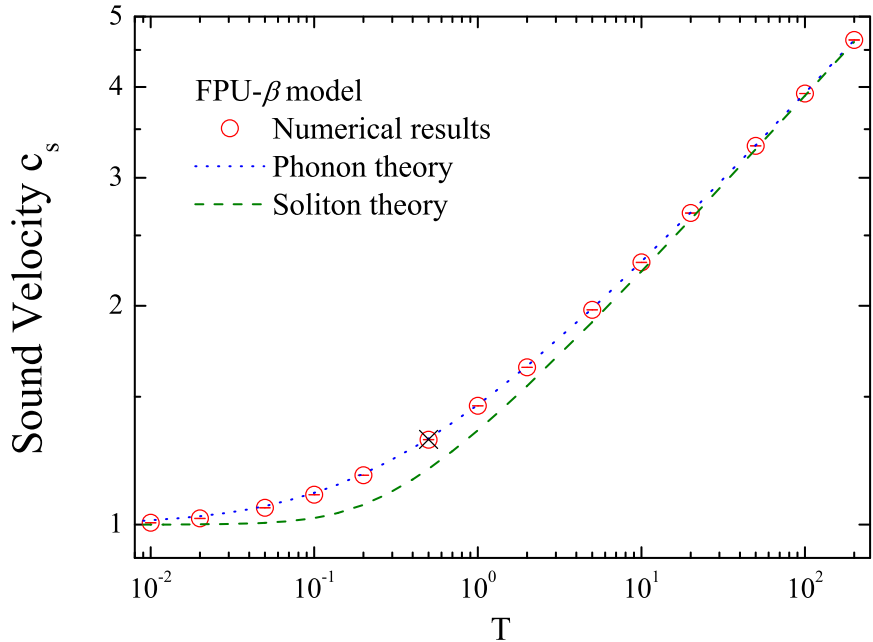}
\vspace{-0.5cm} \caption{\label{fig:cs-fpub} (color online). Sound
velocity $c_s$ as the function of temperature $T$ for FPU-$\beta$
lattice. The circles are the numerical data. The dotted line is the
theoretical prediction from renormalized phonons from Eq.
(\ref{cs-fpub}). The dashed line is the theoretical prediction for
solitons derived from the formula of Eq. (\ref{cs-soliton}) with
$\eta=2.215$. The cross symbol is the numerical data for $T=0.5$
obtained in Ref. \cite{zhaohong2006PRL96}. Adapted from Ref.
\cite{linianbei2010PRL105}.}
\end{figure}

The sound velocity of renormalized phonons can be derived from
$c_s=\frac{\partial{\hat{\omega}_q}}{\partial{q}}|_{q=0}$ where we
have applied the substitution $q=2\pi k/N$ and $\omega_q=2\sin{q/2}$
in the continuous limit. For the well-known FPU-$\beta$ lattice with
$v_2=v_4=1$, the sound velocity of renormalized phonons can be
obtained from Eq. (\ref{ham}) , (\ref{renorm}) and
(\ref{renorm-omega}):
\begin{equation}\label{cs-fpub}
c_s=\sqrt{\alpha}=\sqrt{1+\frac{\int^{\infty}_{0}dx\cdot x^4
e^{-(x^2/2+x^4/4)/T}}{\int^{\infty}_{0}dx\cdot x^2
e^{-(x^2/2+x^4/4)/T}}}
\end{equation}
where dimensionless unit has been used and the Boltzmann constant
$k_{B}$ has been set as unity. The sound velocity depends on the
temperature, or equivalently the strength of nonlinearity. In the
low temperature limit $T\rightarrow 0$, $c_s\rightarrow 1$ as
expected for Harmonic lattice. In the high temperature region
$T>>1$, the sound velocity $c_s\approx 1.22T^{1/4}$ exhibits strong
nonlinear dependence. On the other hand, the soliton theory predicts
a temperature dependent sound velocity as
\begin{equation}\label{cs-soliton}
c^3_s\sqrt{c^2_s-1}=\eta T
\end{equation}
where $\eta$ is a fitting parameter. By tuning $\eta=2.215$, the
sound velocity from soliton theory of Eq. (\ref{cs-soliton})
coincides with the prediction from renormalized phonons of Eq.
(\ref{cs-fpub}) in low and high temperature region. However, these
two predictions deviate from each other in the intermediate
temperature region and can only be verified by accurate numerical
calculations\cite{zhaohong2006PRL96}.

In Fig. \ref{fig:cs-fpub}, the numerically calculated sound velocity
$c_s$ for FPU-$\beta$ lattice has been compared with the prediction
from the theory of renormalized phonons and
solitons\cite{linianbei2010PRL105}, respectively. It can be clearly
seen that the soliton prediction deviates from numerical data in the
intermediate temperature region while the prediction of renormalized
phonons matches perfectly with the numerical data in all temperature
regions. This is a clear evidence that the energy carriers in the
FPU-$\beta$ lattice are the renormalized phonons rather than the
previously believed solitons.

To demonstrate the validity and consistency of the renormalized
phonon formulation, the other three $H_n$ models with $n=3,4,5$ have
been considered with the following Hamiltonian:
\begin{equation}\label{hn-lattice}
H_n=\sum^N_{i=1}\left[\frac{p^2_i}{2}+\frac{\left|x_{i+1}-x_i\right|^n}{n}\right]
\end{equation}
The $H_4$ model is just the high temperature limit of FPU-$\beta$
lattice. In the framework of renormalized phonon formulation, the
sound velocities of $H_n$ model can be expressed in a compact form:
\begin{equation}\label{cs-hn}
c_s=\sqrt{\alpha}=\sqrt{\frac{\int^{\infty}_{0}dx\cdot x^n
e^{-\frac{x^n}{nT}}}{\int^{\infty}_{0}dx\cdot x^2
e^{-\frac{x^n}{nT}}}}
=\sqrt{\frac{\Gamma\left(\frac{n+1}{n}\right)}{\Gamma\left(\frac{3}{n}\right)}}\left(nT\right)^{\frac{1}{2}-\frac{1}{n}}
\end{equation}

\begin{figure}
\includegraphics[width=0.8\columnwidth]{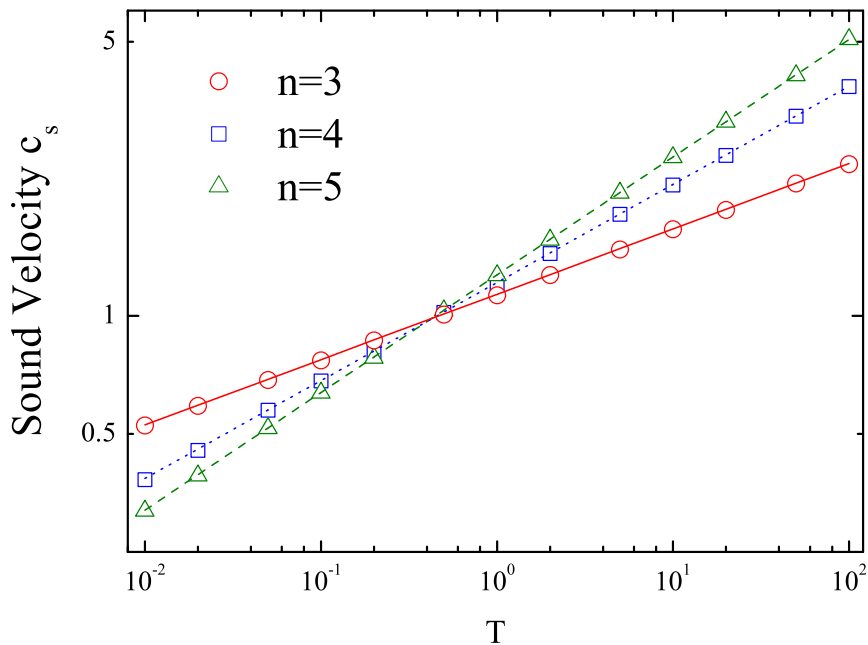}
\vspace{-0.5cm} \caption{\label{fig:cs-hn} (color online). Sound
velocity $c_s$ as the function of temperature $T$ for $H_n$ lattices
with $n=3,4,$ and $5$. The symbols are the numerical data, whereas
the lines are the prediction for renormalized phonons from Eq.
(\ref{cs-hn}). Adapted from Ref. \cite{linianbei2010PRL105}.}
\end{figure}

In Fig. \ref{fig:cs-hn}, the numerically calculated sound velocities
are compared with the predictions for renormalized phonons for $H_n$
lattices with $n=3,4,$ and $5$. Again, the {\it quantitative}
agreements have been found for all three models which we considered.

The above examples are for the lattices without on-site potential.
For the lattices with on-site potential such as $\phi^4$ lattice,
there are numerical results suggesting that the energy carriers are
also renormalized phonons\cite{piazza2009PRB79}.

\section{Effective Phonon Theory}
So far as we know, there is no microscopic transport theory which is
able to explain the heat transport in nonlinear lattices with strong
nonlinearity. The perturbation theory will become invalid as the
nonlinear term cannot be ignored anymore in this case. For example,
no existing theory can even predict the temperature dependence of
thermal conductivities for general nonlinear lattices. A
phenomenological theory possessing such kind of prediction power is
thus highly desirable.

The thermal conductivity for solids is usually described by the
Deybe formula:
\begin{equation}
\kappa=\frac{c}{2\pi}\int^{2\pi}_{0}dqv^2_{q}\tau_{q}
\end{equation}
where $c$ denotes the specific heat, $v_q$ denotes the group
velocity and $\tau_q$ denotes the relaxation time for phonon mode
$q$. The key issue for the Deybe formula is then to determine the
relaxation time $\tau_q$.

For nonlinear lattices, we have shown the evidences that the energy
carriers should be the renormalized phonons. Therefore, the
transport theory should be constructed with the basis of
renormalized phonons. In consideration of the nonlinearity and
anomalous size-dependence of thermal conductivity, we propose the
modified Deybe formula in the framework of renormalized phonons:
\begin{equation}\label{ept-kappa}
\kappa=\frac{c}{2\pi}\int^{2\pi}_{0}dqv^2_{q}\tau_{q}P(q)
\end{equation}
where $P(q)$ is a weight factor with normalization condition
$\frac{1}{2\pi}\int^{2\pi}_{0}dqP(q)=1$, $v_q$ is the group velocity
of renormalized phonons as
$v_q=\partial{\hat{\omega}_q}/\partial{q}$, and $\tau_q$ is the
relaxation time for renormalized phonons which assumes the following
proportionality:
\begin{equation}\label{ept-tau}
\tau_q\propto \frac{1}{\epsilon}\cdot\frac{2\pi}{\hat{\omega}_q}
\end{equation}
where $\hat{\omega}_q$ is the frequency for renormalized phonons and
the dimensionless parameter $\epsilon$ represents the strength of
nonlinearity defined as:
\begin{equation}\label{epsilon}
\epsilon=\frac{\left|\left<E_n\right>\right|}{\left<E_l+E_n\right>},\,\,\,0\leq
\epsilon\leq 1
\end{equation}
where $E_l$ denotes the linear potential energy and $E_n$ denotes
the nonlinear potential energy. The parameter $\epsilon$ describes
the ratio of nonlinear potential energy to the total potential
energy and is independent of mode $q$.

The introduction of the weight factor $P(q)$ is to be consistent
with the anomalous size-dependent thermal conductivities for
lattices without on-site potential such as the FPU-$\beta$ lattice.
If we assume $P(q)\propto 1/q^{\delta}$ in the long wave-length
limit $q\rightarrow 0$, it is straight forward to obtain the
size-dependent $\kappa$ as
\begin{equation}
\kappa\propto\int^{2\pi}_{0^+}dqv^2_q\tau_qP(q)\propto
\int^{2\pi}_{0^+}dq\frac{1}{\omega_q}\cdot\frac{1}{q^{\delta}}\propto
q^{-\delta}\propto N^{\delta}
\end{equation}
the lower integration limit $0^+$ describes the fact that the
lattice size is always finite with length $N$ and $\omega_q\propto
q\propto 1/N$ in the long wave-length limit $q\rightarrow 0$ due to
the finite size effect. The exact value of exponent $\delta$ is
still unresolved and is argued to be $2/5$ from mode-coupling theory
or $1/3$ from hydrodynamic theory\cite{lepri2003PR377}. However,
there are one exception that the momentum conserved one-dimensional
coupled rotator model without on-site potential does exhibit normal
heat conduction behavior\cite{giardina2000PRL84,gendelman2000PRL84}.
Most recently, some numerical works even suggest that the momentum
conserved one-dimensional lattice models with asymmetric
interactions might also follow the Fourier's normal heat conduction
law\cite{zhongyi2012PRE85,lee-dadswellPRE82}.

It should be noticed that the introduction of $P(q)$ simultaneously
explains the normal heat conduction for lattices with on-site
potential. The non-zero renormalization coefficient $\gamma$ shifts
the renormalized phonon frequency from zero to
$\hat{\omega}_q=\sqrt{\gamma}$ in the long wave-length limit
$q\rightarrow 0$. The integral in last equation becomes
\begin{equation}
\kappa\propto\int^{2\pi}_{0^+}dq\frac{1}{\hat{\omega}_q}\cdot\frac{1}{q^{\delta}}
<\frac{1}{\sqrt{\gamma}}\int^{2\pi}_{0^+}dq\frac{1}{q^{\delta}}
\end{equation}
One should notice that $\delta<1$ due to the constriction of
normalization condition for $P(q)$ as
$\frac{1}{2\pi}\int^{2\pi}_{0}dqP(q)=1$. The integral is
non-divergent and the thermal conductivities for lattices with
on-site potential is finite and obeys Fourier's heat conduction law
which is consistent with previous
results\cite{hubambi1998PRE57,hubambi2000PRE61,aoki2000PLA265}.

In this paper, we will only focus on the temperature dependent
thermal conductivities for nonlinear lattices without on-site
potential. The renormalization coefficient $\gamma=0$ and the
renormalization phonon frequency
$\hat{\omega}_q=\sqrt{\alpha}\omega_q$ where temperature dependent
part $\alpha$ and mode dependent part $\omega_q$ are separate. From
Eq. (\ref{ept-kappa}) and (\ref{ept-tau}), the temperature
dependence of thermal conductivity can be fully described by the
simple formulation
\begin{equation}\label{kappa-fpu-t}
\kappa(T)\propto\frac{\sqrt{\alpha}}{\epsilon}\propto
\frac{c_s}{\epsilon}
\end{equation}
where the renormalization coefficient $\alpha$ (or the sound
velocity $c_s$) and nonlinearity strength $\epsilon$ are fully
determined by the lattice Hamiltonian. One should also notice that
these two parameters only depend on temperature and are mode
independent. In another word, the size- and temperature-dependences
of the thermal conductivities are separate, which enables us to
discuss the temperature-dependence of the thermal conductivities
alone without worrying about the annoying size-dependence. This
effect has been verified by our numerical simulations (not shown
here).

\begin{figure}
\includegraphics[width=0.8\columnwidth]{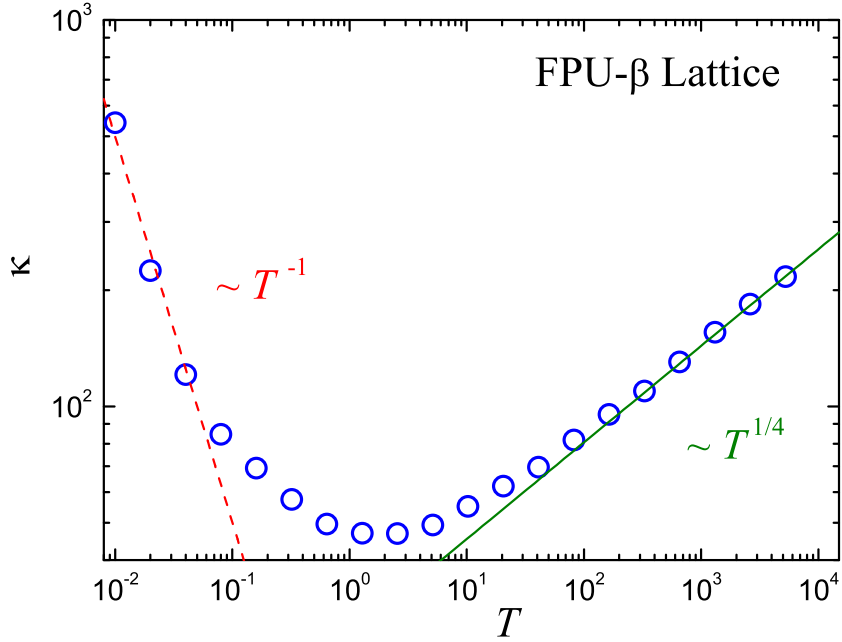}
\vspace{-0.5cm} \caption{\label{fig:kappa-fpu} (color online).
Thermal conductivity $\kappa$ as the function of temperature $T$ for
FPU-$\beta$ lattice. The lattice size is set as $N=1000$. The
circles are the numerical data from NEMD simulations.}
\end{figure}

For the FPU-$\beta$ lattice with Hamiltonian of Eq. (\ref{fpu-ham}),
the temperature dependence of $c_s$ and $\epsilon$ exhibits scaling
relations at both low and high temperature limits. As we have
discussed above, the sound velocity $c_s$ shows the following
temperature dependence:
\begin{equation}\label{cs-fpu-t}
c_s \propto \mbox{const}, \,T<<1;\,\,\,c_s\propto T^{1/4}, \,T>>1
\end{equation}

From definition of Eq. (\ref{epsilon}), the nonlinearity strength
$\epsilon$ for FPU-$\beta$ lattice can be expressed as
\begin{equation}\label{epsilon-fpu}
\epsilon=\frac{\left<\sum_{i}(x_{i+1}-x_i)^4/4\right>}{\left<\sum_{i}(x_{i+1}-x_i)^2/2+\sum_{i}(x_{i+1}-x_i)^4/4\right>}
\end{equation}
At low temperature limit, the quartic potential term is a small term
comparing to the quadratic potential term. The whole system can be
described by a harmonic lattice plus a small perturbation. From
equipartition theorem for harmonic lattice, we thus have
$\left<\sum_i(x_{i+1}-x_{i})^2\right>\approx Nk_{B}T$ for the
FPU-$\beta$ lattice at low temperature limit. This also gives rise
to the scaling relation for the quartic potential term, i.e.
$\left<\sum_i(x_{i+1}-x_{i})^4\right>\propto T^2$. Thus the
nonlinearity strength $\epsilon$ follows
\begin{equation}
\epsilon\approx\frac{\left<\sum_{i}(x_{i+1}-x_i)^4/4\right>}{\left<\sum_{i}(x_{i+1}-x_i)^2/2\right>}\propto
T
\end{equation}

On the other hand, the quadratic potential term is a small term at
high temperature limit and the nonlinearity strength $\epsilon$
approaches to the upper limit $1$. The entire temperature dependence
of $\epsilon$ can be summarized as:
\begin{equation}\label{epsilon-fpu-t}
\epsilon\propto T,\,T<<1;\,\,\,\epsilon \propto \mbox{const},\,T>>1
\end{equation}

From Eq. (\ref{kappa-fpu-t}), (\ref{cs-fpu-t}) and
(\ref{epsilon-fpu-t}), we can derive the temperature dependence of
$\kappa$ for FPU-$\beta$ lattice at both low and high temperature:
\begin{equation}\label{kappa-fpu-all-t}
\kappa(T)\propto \frac{1}{T},\,T<<1;\,\,\,\kappa(T)\propto
T^{1/4},\,T>>1
\end{equation}

In Fig. \ref{fig:kappa-fpu}, the numerical results of thermal
conductivities $\kappa$ calculated by Non-Equilibrium Molecular
Dynamics (NEMD) simulations are plotted for FPU-$\beta$ lattices. It
can be seen that the scaling of the thermal conductivities follows
the relations $\kappa\propto 1/T$ at low temperature limit and
$\kappa\propto T^{1/4}$ at high temperature limit, respectively. The
temperature dependence of $\kappa$ for FPU-$\beta$ lattice can thus
be {\it consistently} explained by our effective phonon theory at
both low and high temperature limits.

\begin{figure}
\includegraphics[width=0.8\columnwidth]{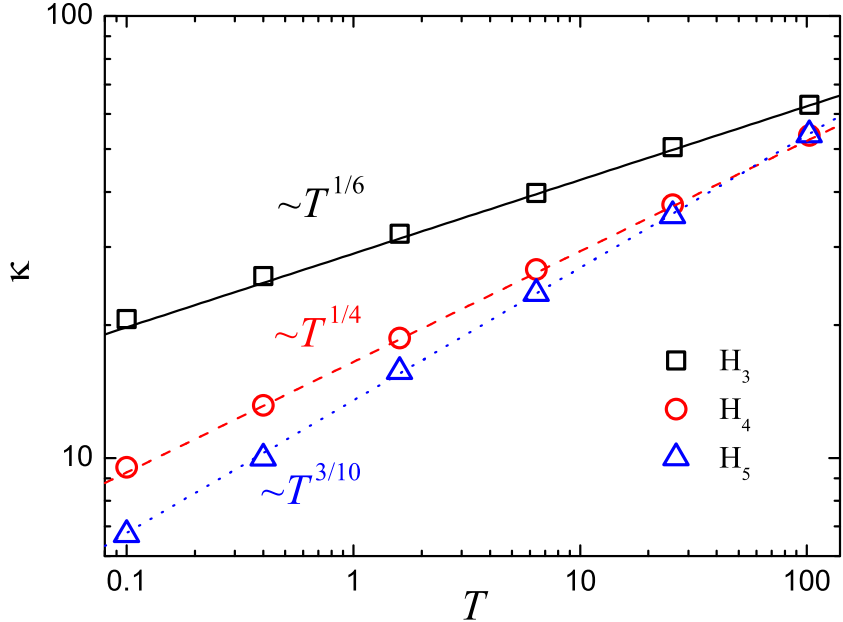}
\vspace{-0.5cm} \caption{\label{fig:kappa-hn} (color online).
Thermal conductivity $\kappa$ as the function of temperature $T$ for
$H_n$ lattices with $n=3,4$ and $5$. The lattice sizes are all set
as $N=500$. The symbols are the numerical data from NEMD
simulations.}
\end{figure}

To further verify the predictions from effective phonon theory, we
also consider the $H_n$ lattices with Hamiltonian of Eq.
(\ref{hn-lattice}). The fact that the potential energy of these
$H_n$ lattices is fully nonlinear gives rise to the special property
for the nonlinearity strength as $\epsilon=1$. From Eq.
(\ref{cs-hn}), the temperature dependence of the sound velocity
$c_s$ follows $c_s\propto T^{1/2-1/n}$. The thermal conductivities
$\kappa$ for $H_n$ lattices can be derived as:
\begin{equation}
\kappa\propto c_s\propto T^{\frac{1}{2}-\frac{1}{n}}
\end{equation}
Therefore, from effective phonon theory, $\kappa\propto T^{1/6},
T^{1/4}$ and $T^{3/10}$ for $H_3$, $H_4$ and $H_5$ lattices,
respectively. From Fig. \ref{fig:kappa-hn}, the numerical results of
thermal conductivities for all these $H_n$ lattices are plotted and
compared with the theoretical predictions from effective phonon
theory. Perfect agreements between theory and numerical results have
been found for all the $H_n$ lattices at temperature regions over
three orders of magnitudes.

\section{Conclusions}
With this short review, we have briefly introduced the heat
conduction of one-dimensional nonlinear lattices and the role of
renormalized phonons and effective phonon theory. Although there are
more than one type of excitation modes in the nonlinear lattices,
detailed analysis on the temperature dependence of sound velocities
reveals that the energy carriers responsible for the heat conduction
should be the renormalized phonons.

Within the framework of renormalized phonons, a phenomenological
theory, effective phonon theory, has been developed to explain the
heat transport behaviors for general one-dimensional nonlinear
lattices. For lattices without on-site potential, the effective
phonon theory can {\it quantitatively} and {\it consistently}
predict the scaling exponents of temperature-dependent thermal
conductivities. In particular, for the FPU-$\beta$ lattices, the
predictions from effective phonon theory are verified by NEMD
simulations at both low and high temperature limits. For the special
class of $H_n$ lattices, the predictions from effective phonon
theory are in perfect agreements with the obtained numerical
results.

Finally, we need to point out that our work is closely related to
the emerging field of phononics which focuses on the manipulation of
heat flow and information processing with
phonons\cite{linianbei2012RMP}. For a two-segment thermal diode, the
rectification of heat flow with the inversion of a temperature
gradient is realized by the temperature-modulated overlap of the
phonon spectra of two segments. It will be crucial to know the
temperature dependence of the phonon spectrum in advance. Our
analysis of the renormalized phonons due to nonlinearity or
temperature enables us to provide the key information for the basic
design of thermal diodes. On the other hand, our effective phonon
theory dealing with the analysis of temperature-dependent thermal
conductivities can be also very helpful for the design of novel
phononics devices. The key element of phononics devices is the
thermal transistor which relies on an effect of negative
differential thermal resistance (NDTR). And this NDTR effect is a
consequence of the system's very steep temperature-dependent thermal
conductivity, e.g. $\kappa \propto T^{\pm \eta}$ with $\eta>1$ at
least. Therefore, we hope our work about renormalized phonons and
effective phonon theory can provide solid theoretical support for
the ongoing research filed of phononics\cite{linianbei2012RMP}.

\section{Acknowledgements}
Parts of the work were carried out at National Supercomputer Center
in Tianjin, and the calculations were performed on TianHe-1£¨A£©.
This work has been supported by the startup fund from Tongji
University (N. L. and B. L.).

\end{document}